# Low Temperature Specific Heat of Doped SrTiO$_3$: Doping Dependence of the Effective Mass and Kadowaki-Woods Scaling Violation

E. McCalla[1,2], M.N. Gastiasoro[3], G. Cassuto[1], R.M. Fernandes[3] and C. Leighton[1]

[1]*Department of Chemical Engineering and Materials Science, University of Minnesota, Minneapolis, Minnesota, USA*
[2] *Department of Chemistry, McGill University, Montreal, Canada*
[3]*School of Physics and Astronomy, University of Minnesota, Minneapolis, Minnesota, USA*

**Abstract:** We report wide-doping-range (8 × 10$^{17}$ to 4 × 10$^{20}$ cm$^{-3}$ Hall electron density) low temperature specific heat measurements on single crystal SrTiO$_3$:Nb, correlated with electronic transport data and tight-binding modeling. Lattice dynamic contributions to specific heat are shown to be well understood, albeit with unusual sensitivity to doping, likely related to the behavior of soft modes. Electronic contributions to specific heat provide effective masses that increase substantially, from 1.8 to 4.8$m_e$, across the two SrTiO$_3$ Lifshitz transitions. It is shown that this behavior can be quantitatively reconciled with quantum oscillation data and calculated band structure, establishing a remarkably doping-independent mass enhancement factor of 2.0. Most importantly, with the doping-dependent $T^2$ resistivity prefactor and Sommerfeld coefficient known, Kadowaki-Woods scaling has been tested over the entire doping range probed. Despite classic Fermi liquid behavior in electronic specific heat, standard Kadowaki-Woods scaling is dramatically violated, highlighting the need for new theoretical descriptions of $T^2$ resistivity in SrTiO$_3$.

**Section:** M9-A: Other electronic materials

**PhySH**: Research areas: Electrical properties, electronic structure
Physical systems: Complex oxides, doped semiconductors
Techniques: Specific heat measurements, transport techniques

**Corresponding author**: leighton@umn.edu

Few materials have posed such challenges to condensed matter physics as the perovskite oxide $SrTiO_3$. From the structural and lattice dynamic perspective, this material has revealed an extraordinary low temperature ($T$) quantum paraelectric state where ferroelectricity is suppressed by quantum fluctuations [1-5], in addition to a second-order antiferrodisplacive structural transformation at 105 K [5-9]. The former is associated with $T \rightarrow 0$ softening of a zero wavevector transverse phonon [1-5], whereas the latter is related to softening of a zone-boundary octahedral rotation mode [5-9]. Quantum paraelectricity also leads to interesting physics when $SrTiO_3$ is doped. $n$-doping with $Nb^{5+}$ and $La^{3+}$ (for $Ti^{4+}$ and $Sr^{2+}$) has been explored, along with oxygen vacancy doping [10-14]. Due to the high dielectric constant (>10,000 at low $T$ [1,3,4]), donors in $SrTiO_3$ have unusually large Bohr radii (~600 nm), vanishing ionization energies, and highly screened ionized scattering potentials [10-14]. Donor wavefunctions thus overlap at very low electron density ($n$), generating a remarkable low-density high-mobility metallic state [10-14]. Metallic transport has been claimed to $n < 10^{16}$ cm$^{-3}$ [14] in fact, with a well-defined Fermi surface down to ~$10^{17}$ cm$^{-3}$ [15,16].

The very small Fermi surface of this dilute metal has been studied by Shubnikov-de Haas (SdH) oscillations [15-18], Angle-Resolved Photoemission Spectroscopy (ARPES) [19], and Density Functional Theory (DFT) [20-22]. In the low $T$ tetragonal state, the $Ti^{4+}$ $t_{2g}$ states at the conduction band minimum are split by tetragonality (by ~2-5 meV) and spin-orbit interactions (by ~12-30 meV) [15-22]. Two Lifshitz transitions thus occur $vs.$ $n$ in $SrTiO_3$, corresponding to first occupation of the tetragonality-split band (at $n_{c1} \approx 1.5 \times 10^{18}$ cm$^{-3}$) and the spin-orbit-split band (at $n_{c2} \approx 2 \times 10^{19}$ cm$^{-3}$) [15-22]. The three bands have varying anisotropies and dispersions, leading to effective mass ($m^*$) that varies from ~$1.5m_e$ ($m_e$ is the free electron mass) to 4-6 $m_e$ as the Lifshitz transitions are crossed [16]. While the agreement on band splittings and masses among SdH measurements, SdH and ARPES, and experiment and theory is generally reasonable, significant uncertainties remain, particularly at high $n$.

Superconductivity leads to further interest in SrTiO$_3$. This was in fact the first oxide superconductor discovered, the first semiconductor known to superconduct, and the first example of a superconducting "dome" [11,23]. Superconductivity also occurs to extraordinarily low $n$ (3 × 10$^{17}$ cm$^{-3}$), with subtle interplay with band filling [16,24], leading to several proposals for novel pairing mechanisms [25-29]. Recently, normal state transport in doped SrTiO$_3$ has also come under intense scrutiny [21,30-34]. In particular, below 60-100 K SrTiO$_3$ has been found to exhibit the $T^2$ resistivity ($\rho$) often taken as evidence of Fermi liquid behavior [21,30-34]. This is a puzzling observation, however. At $n < n_{c1}$, for example, the tiny Fermi surface, low Fermi temperature, and single filled electron reservoir appear to rule out the scattering processes (*e.g.*, umklapp) typically needed to generate resistivity of the form $\rho \propto \rho_0 + AT^2$ [32]. The electron scattering rate is also independent of $n$ over four orders of magnitude (or, equivalently, $A \propto 1/n$), which has been stated to be at odds with theory [33,34]. These observations, among others, have raised doubts over the Fermi liquid nature of the metallic state in this foundational oxide.

In principle, heat capacity ($C_P$) measurements have the potential to greatly elucidate much of the above. Lattice dynamic contributions to $C_P$, for example, could probe the complex low $T$ evolution of phonon modes in SrTiO$_3$. Unfortunately, even basic parameters such as the Debye temperature, $\theta_D$, are remarkably scattered in the SrTiO$_3$ literature, as discussed in Supplemental Material Section A (Table SI) [35] [30,31,36-42]. Moreover, electronic contributions to $C_P$ could probe: the existence of a well-defined $\gamma T$ contribution (where $\gamma$ is the Sommerfeld coefficient), as expected in a Fermi liquid; the density-of-states effective mass ($m^*_{DOS}$) *vs.* $n$ (for comparison to SdH, DFT, etc.); and Kadowaki-Woods scaling. The latter refers to the well-known linear scaling between $A$ and $\gamma^2$ that is empirically established in Fermi liquids such as transition metals and heavy fermion compounds [43,44]. Again, however, the literature on $\gamma(n)$ (and thus $m^*_{DOS}(n)$) in SrTiO$_3$ is highly inconsistent, as shown in Supplemental Material Section A, Fig. S1 [35]. The use of polycrystalline [36-38,41] and potentially inhomogeneous [36,41] samples, impurity-related Schottky anomalies [38,39], and the limited doping ranges studied [30,31,36-42], all likely contribute to this inconsistency. Intriguingly, however, while the number of data points is very limited, existing data on $A$ and $\gamma$ do indicate potential violation of Kadowaki-Woods scaling in La-doped SrTiO$_3$ [30,31].

Here, we rectify this situation through a wide $n$ range ($8 \times 10^{17}$ to $4 \times 10^{20}$ cm$^{-3}$) low temperature $C_P(T)$ study of thoroughly-characterized SrTiO$_3$:Nb single crystals, correlated with $\rho(T)$ measurements and tight-binding modeling. It is shown that lattice dynamic contributions to $C_P$ can be understood, albeit with unusual sensitivity to doping, potentially related to incipient ferroelectricity. A well-defined Fermi-liquid-like $\gamma T$ contribution to $C_P$ is then isolated, supported by theory, providing detailed $\gamma(n)$ data for comparison to $A(n)$. The $n$-dependence of $m^*_{DOS}$ is thus significantly clarified, and reconciled with SdH measurements and band structure calculations, establishing a completely *n-independent* mass enhancement factor of 2.0. Most importantly, despite the Fermi-liquid-like electronic $C_P$, $A$ is found to *decrease* by two orders of magnitude with increasing $\gamma$, leading to striking violation of standard Kadowaki-Woods scaling, with deep implications for the origin of the $T^2$ resistivity.

The Nb-doped single crystals studied here are some of the same ones used in prior work, and have been characterized by X-ray diffraction [14], impurity analysis [45], and resistivity, Hall effect, mobility ($\mu$), and magnetoresistance measurements [14]. Nb content ($x$ in SrTi$_{1-}$ $_x$Nb$_x$O$_3$), $n$(300 K), $\rho$(300 K), and $\mu$(4 K) are shown in Table SII of Supplemental Material Section B [35], which also includes a discussion of methods. Briefly, short-pulse relaxation calorimetry was used, with great attention paid to errors associated with thermal coupling and sample-to-addenda $C_P$ ratios [46]. Fig. 1(a) first shows wide $T$ range $C_P(T)$ for these crystals. Aside from small anomalies around the structural transformation at 105-130 K (with doping dependence studied in prior work [9]), $C_P(T)$ is qualitatively as expected, with only minor apparent $x$ dependence. Fig. 1(b) shows typical analysis of low $T$ data (1.8 to 9 K), where $C_P/T$ is plotted $vs.$ $T^2$ to test the relation $C_p(T) = \beta T^3 + \gamma T$. Here, $\beta T^3$ is the Debye lattice term (where $\beta = 234Nk_B/\theta_D$, $N$ is the number of ions/mole, and $k_B$ is Boltzmann's constant), and $\gamma T$ captures electronic contributions. In a single band free electron model, $\gamma = m^*_{DOS} n^{1/3}(k_B/\hbar)^2(\pi/3)^{2/3}$, where $\hbar = h/2\pi$ and $h$ is Planck's constant. As illustrated by the solid line fits in the inset to Fig. 1(b), this form describes the data very well up to 3-4 K, with an intercept, $\gamma$, that increases with $n$. As shown in the main panel, however, at higher $T$, up to 9 K, upward curvature is apparent in $C_P/T$ $vs.$ $T^2$. As is often required, we thus add a next-order Debye term in $C_P(T) = \beta T^3 + \alpha T^5 + \gamma T$, resulting in the good fits shown in Fig. 1(b). The resulting $\theta_D$, $\alpha$, and $\gamma$ are plotted in Fig. 2.

We discuss first the lattice dynamic contributions to $C_P(T)$, returning below to electronic contributions. While the undoped SrTiO$_3$ $\theta_D$ of 515 ± 20 K (Fig. 2(a)) is larger than many of the early, widely scattered values (see Supplemental Material Section A [35]), it is in excellent agreement with recent single crystal work [40]. Surprisingly, however, $\theta_D$ is sensitive to even light doping. As shown in Fig. 2(a), $\theta_D$ increases to 570-590 K at $x$ = 1%, where it plateaus. This increase is above uncertainty, and is in fact readily apparent in Fig. 1(b), where the slope clearly decreases with doping. Interestingly, this increase in $\theta_D$, and upward curvature in $C_P(T)$ $vs.$ $T^2$, were also noted by Ahrens $et$ $al$. [40], who rejected the possibility of such sensitivity to doping and thefore fixed $\theta_D$ at its undoped value. In our case, $\theta_D$ is independent of whether fitting is performed over the range in the inset to Fig. 1(b), with no $T^5$ term, or the range in the main panel, with $T^5$ included. We thus have high confidence in the $\theta_D(x)$ in Fig. 2(a), whose behavior is mirrored in $\alpha(x)$ (Fig. 2(b)). Note that in all cases the $\alpha T^5$ contribution to $C_P$ is indeed substantially smaller than $\beta T^3$, as expected.

We contend that established trends in soft mode frequencies in SrTiO$_3$ provide potential explanations for Fig. 2(a,b). It is known, for example, that the antiferrodisplacive transformation temperature shifts from 105 to 130 K in this $x$ range [9], which could increase the low $T$ frequencies of the corresponding modes, thereby increasing $\theta_D$. The transformation temperature shift is linear in $x$, however, which is difficult to reconcile with the sharp increase at low $x$ seen in Fig. 2(a,b). Alternatively, the $T \rightarrow 0$ softening of the ferroelectric mode in SrTiO$_3$ should also weaken with doping (*i.e.*, the frequency should increase), due to screening of inter-dipole interactions. This is analogous to the situation in Sr$_{1-x}$Ca$_x$TiO$_{3-\delta}$, where increased $n$ suppresses the cubic-to-tetragonal "ferroelectric" transition [47]. Simple estimates indicate that, taking into account the expected decrease in dielectric constant with doping, the Thomas-Fermi screening length could indeed approach the Ti-O-Ti distance at low $n$ in SrTiO$_3$, potentially explaining Fig. 2(a).

A final interesting point on lattice dynamics is highlighted in Fig. 1(c), where $C_P/T^3$ is plotted *vs. T*, to higher $T$ than in Fig. 1(b). Such plots expose deviations from Debye behavior (dashed line), which are apparent in SrTiO$_3$ as a peak in excess $C_P$ around 30 K. This is a known phenomeon in perovskite oxides, thought to occur due to excess lattice $C_p$ associated with the first maximum in the phonon density-of-states [40,48]. The solid line in Fig. 1(c) is a fit to the Einstein expression, $C_P(T) = 3Rw(\theta_E/2T)^2\sinh^{-2}(\theta_E/2T)$, where $R$ is the gas constant, $w$ is a weight factor, and $\theta_E$ is the Einstein temperature ($\hbar\omega_0/k_B$, where $\omega_0$ is the phonon frequency). This describes the excess $C_P$ well, with $\hbar\omega_0$ = 12.9 meV, in good agreement with prior work [40] and the first peak in the SrTiO$_3$ phonon density-of-states [49]. The peak in Fig. 1(c) is $x$-independent in this range, as expected for these particular phonons, which are not soft modes.

We now turn to electronic contributions to $C_P(T)$. As noted above, analysis of Fig. 1(b) suggests that these may be captured with the standard $\gamma T$ Fermi liquid form. It is not, however, *a priori* clear that such analysis is even valid in this $T$ range; the Fermi energy and temperature are as low as 1.7 meV and 16 K, and Fermi liquid behavior has been questioned. To inform our analysis we thus used a tight-binding fit to the DFT-calculated structure of the conduction band minimum in SrTiO$_3$ [21], as shown in Fig. 3(a). Shown here are the three $t_{2g}$-derived bands, the tetragonal and spin-orbit splittings, and the critical densities for the two Lifshitz transitions, $n_{c1}$, and $n_{c2}$. The electronic specific heat, $C_V^{elec}(T)$, was then computed from a self-consistent calculation of the chemical potential $\mu(T)$ and eigenstates $E_{ik}(T)$ of each band $i$, for a given $n$, using a 120×120×120 three-dimensional $k$ grid. $C_V^{elec}(T) = T\frac{\partial S}{\partial T}$ is then calculated from the entropy, $S(T) = -2k_B \sum_{ik}\{f(E_{ik})\log[f(E_{ik})] + f(-E_{ik})\log[f(-E_{ik})]\}$. The results are shown in Fig. 1(d), where $C_V/T$ is plotted *vs*. $T$. While deviations from linearity are present (compare the data to the horizontal dashed lines), particularly at low $n$, even at 7.8 × 10$^{17}$ cm$^{-3}$ these occur only above ~15 K, validating our analysis of Fig. 1(b). The thus extracted $\gamma(n)$ are shown in Fig. 2(c), along with the calculated $\gamma$, which, for reasons clarified below, are multiplied by a factor of 2.0. These values match well the experimental data.

More detailed analysis is provided in Fig. 3(b), which plots $m^*_{DOS}(n)$ (solid black points) extracted from $\gamma(n) = m^*_{DOS}(n)n^{1/3}(k_B/\hbar)^2(\pi/3)^{2/3}$, using the measured Hall densities. We find $m^*_{DOS} \approx 1.8 m_e$ below $n_{c1}$, increasing to $\sim 4 m_e$ at $n_{c2}$, before plateauing at 4.5-5$m_e$. Note that the systematics are greatly improved over prior results from $C_P(T)$, the data also extending to lower $n$ (see Supplemental Material Section A [35]). Quantitative consistency with SdH measurements can in fact be demonstrated. To this end, the blue, red, and green dashed lines in Fig. 3(b) approximate the SdH-determined $m^*$ in the 1st, 2nd, and 3rd bands in Fig. 3(a) [16]. Starting at $n < n_{c1}$, where only one band is occupied, we find good agreement between $m^*_{DOS}$ from $C_P$ and $m^*$ from SdH. For $n_{c1} < n < n_{c2}$, the SdH $m^*$ in band 2 then stays constant, while the $m^*$ in band 1 increases sharply, as expected from Fig. 3(a). Importantly, $m^*_{DOS}$ reflects a weighted sum of these SdH masses. Specifically, since $\gamma(n) \propto n^{1/3} m^*$, we write (see Supplemental Material Section C [35] for justification) $m^*_{DOS} = \sum_{i=1}^{3} n_i^{1/3} m_i^* / (\sum_{i=1}^{3} n_i)^{1/3}$, where the $m^*_i$ and $n_i$ are SdH masses and electron densities. SdH data then predict $m^*_{DOS}$ should increase to 4.9$m_e$ at $n_{c2}$, in reasonable agreement with our extracted $m^*_{DOS}$. Finally, at $n > n_{c2}$, SdH results become sparse, especially for band 1. $m^*$ values for bands 2 and 3 are available at $1.5 \times 10^{20}$ cm$^{-3}$, however [16], and can be supplemented with a measurement of the ARPES heavy band mass of 7$m_e$ [19] to predict $m^*_{DOS}$ = 5.2$m_e$. This is again in good agreement with our $m^*_{DOS}$, as well as a single $m^*_{DOS}$ point from the work of Lin *et al.* [42]. We thus conclude *quantitative* agreement between SdH measurements and electronic contributions to $C_P(T)$ for the filling-dependent $m^*$ in SrTiO$_3$.

Also plotted in Fig. 3(b) (red circles) are theoretical $m^*_{DOS}(n)$ values from the tight-binding modeling shown in Fig. 3(a). Remarkably, excellent agreement with experiment is obtained simply by multiplying by a constant factor of 2.0. A mass enhancement factor of ~2 in SrTiO$_3$ has been deduced before at certain $n$ [18,21], but is shown here to be completely $n$-independent (see also Fig. 2(c)). In addition to agreement with SdH measurements, we thus conclude quantitative reconciliation with the calculated band structure, with a mass enhancement factor of 2.0. As in prior work, we attribute the modest mass enhancement to effects such as electron-phonon interaction [18,21] or electronic correlations. Whatever the origin is, it is apparently unaffected by doping in the range studied here. Significantly, $n$-independent electron-phonon mass enhancement would also suggest that the $SrTiO_3$ superconducting dome is not caused by variations in pairing interaction.

Finally, with Fermi-liquid-like electronic $C_P$ established, and a detailed wide-range $\gamma(n)$ available, we turn to Kadowaki-Woods scaling. Earlier $\rho(T)$ measurements were supplemented with additional data, and tested for $T^2$ behavior. As shown in Supplemental Material Section D [35] (Figs. S2, S3), $\rho \propto \rho_0 + AT^2$ indeed holds to reasonable confidence below 50-110 K (dependent on $n$), albeit with some deviations at the lowest $T$. As shown in Fig. S3, $T^2$ evolves toward $T^3$ at higher $T$ [50], before the exponent falls again. Fig. S4 shows that the extracted $A$ are in good agreement with prior work [21,30-34], following $A \propto 1/n$ over four orders of magnitude in $n$. Kadowaki-Woods scaling is then tested in Fig. 3(c), which plots $A$ vs. $\gamma^2$ (on a log-log plot), with $n$ as an implicit variable. The empirical $A/\gamma^2 = C$ behavior (where $C$ is a materials-class-specific positive constant) [43,44], shown in Fig. 3(c) for transition metals and heavy fermion compounds (dashed lines), is seen to be *qualitatively* violated in doped $SrTiO_3$. Specifically, our data on $SrTiO_3$:Nb (large black circles) reveal anomalously large $A$ at low $\gamma$, *decreasing* by a factor of 100 as $\gamma^2$ increases by a factor of 400. This trend is qualitatively consistent with the data of Okuda *et al.* [30] on $SrTiO_3$:La (green diamonds), although that data set is sparser, and, as already noted (Supplemental Material Section A, Fig. S1 [35]), differs significantly in terms of the values of $\gamma$, and thus $m^*$. Intriguingly, the decrease in $A$ with $\gamma$ in doped $SrTiO_3$ reverts to the typical Kadowaki-Woods scaling in Sr-doped $LaTiO_3$, *i.e.*, at the La-rich end of the $Sr_{1-x}La_xTiO_3$ series (orange squares) [51]. Kadowaki-Woods scaling is thus obeyed as the Mott insulator

LaTiO$_3$ is hole doped, where electronic correlations are strong, but is qualitatively violated at lower $x$.

A number of approaches have been explored in the literature to account for charge carrier density effects in the Kadowaki-Woods ratio [*e.g.*, 52,53], and we thus attempted to implement those for SrTiO$_3$:Nb. As shown in Fig. S5 (Supplemental Material Section E [35]), the modified Kadowaki-Woods scaling of Jacko *et al.* [52], designed to account for doping and dimensionality effects, is also violated in doped SrTiO$_3$. We note, however, that the approach of Hussey [53] to account for doping effects in SrTiO$_3$ is successful, both for prior data in the Sr$_{1-x}$La$_x$TiO$_3$ system [53], and the current data on SrTiO$_3$:Nb (see Fig. S6, Supplemental Material Section E [35]). Nevertheless, despite clear Fermi-liquid-like behavior in electronic $C_P$, and simply rationalized behavior of $\gamma(n)$, the doping-dependent $T^2$ resistivity prefactor in SrTiO$_3$ appears to differ substantially from simple Fermi liquid expectations. As noted above, it is in fact not even clear why $T^2$ resistivity occurs when umklapp processes appear impossible. This adherence to Fermi liquid behavior for thermodynamic properties, but clear deviation for transport, suggests a potentially atypical origin of the $T^2$ resistivity. Additional theoretical work is clearly needed, including exploring potential explanations beyond electron-electron interactions.

In summary, $C_P(T)$ measurements on single crystal SrTiO$_3$ have been performed over a wide doping range. We conclude: (*i*) that lattice dynamic contributions can be understood, albeit with unusual doping sensitivity, likely related to doping evolution of soft modes; (*ii*) that the extracted $m^*_{DOS}$ can be quantitatively reconciled with SdH measurements and calculated band structure, yielding an $n$-independent mass enhancement factor of 2.0; and (*iii*) that standard Kadowaki-Woods scaling is dramatically violated, despite the Fermi-liquid-like electronic specific heat. These results have deep implications for the origin of the puzzling $T^2$ resistivity in SrTiO$_3$.

**Acknowledgments:** Work supported by the US Department of Energy through the University of Minnesota Center for Quantum Materials under DE-SC-0016371. E.M. thanks the Fonds de Recherche du Québec, Nature et Technologies, and the Natural Science and Engineering Research Council of Canada for fellowship support.

**References**

1. J.H. Barrett, Phys. Rev. **86**, 118 (1952).

2. P.A. Fleury and J.M. Worlock, Phys. Rev. **174**, 613 (1968).

3. H. Uwe and T. Sakudo, Phys. Rev. B. **13**, 271 (1976).

4. K.A. Muller and H. Burkard, Phys. Rev. B. **19**, 3593 (1979).

5. J.F. Scott, Rev. Mod. Phys. **46**, 83 (1974).

6. G. Shirane, Rev. Mod. Phys. **46**, 437 (1974).

7. F.W. Lytle, J. Appl. Phys. **35**, 2212 (1964).

8. P.A. Fleury, J.F. Scott and J.M. Worlock, Phys. Rev. Lett. **21**, 16 (1968)

9. E. McCalla, J. Walter and C. Leighton, Chem. Mater. **28**, 7973 (2016).

10. H.P.R. Frederikse, W.R. Thurber and W.R. Hosler, Phys. Rev. **134**, A442 (1964).

11. C.S. Koonce, M.L. Cohen, J.F. Schooley, W.R. Hosler and E.R. Pfeiffer, Phys. Rev. **163**, 380 (1967).

12. O.N. Tufte and P.W. Chapman, Phys. Rev. **155**, 796 (1967).

13. C. Lee, J. Yahia and J.L. Brebner, Phys. Rev. B **3**, 2525 (1971).

14. A. Spinelli, M.A. Torija, C. Liu, C. Jan and C. Leighton, Phys. Rev. B. **81**, 155110 (2010).

15. X. Lin, Z. Zhu, B. Fauque and K. Behnia, Phys. Rev. X. **3**, 021002 (2013).

16. X. Lin, G. Bridoux, A. Gourgout, G. Seyfarth, S. Kramer, M. Nardone, B. Fauque and K. Behnia, Phys. Rev. Lett. **112**, 207002 (2014).

17. H. Uwe, R. Yoshizaki, T. Sakudo, A. Izumi and T. Uzumaki, Jpn. J. Appl. Phys. **24**, 335 (1985).

18. S.J. Allen, B. Jalan, S. Lee, D.G. Ouellette, G. Khalsa, J. Jaroszynski, S. Stemmer and A.H. MacDonald, Phys. Rev. B **88**, 045114 (2013).

19. Y.J. Chang, A. Bostwick, Y.S. Kim, K. Horn and E. Rotenberg, Phys. Rev. B. **81**, 235109 (2010).

20. L.F. Mattheiss, Phys. Rev. B. **6**, 4740 (1972).

21. D. Van der Marel, J.L.M. van Mechelen and I.I. Mazin, Phys. Rev. B. **84**, 205111 (2011).

22. A. Janotti, D. Steiauf and C.G. Van de Walle, Phys. Rev. B. **84**, 201304 (2011).

23. J.F. Schooley, W.R. Hosier, E. Ambler, J.G. Becker, M.L. Cohen and C.S. Koonce, Phys. Rev. Lett. **14**, 305 (1965).

24. X. Lin, Z. Zhu, B. Fauque and K. Behnia, Phys. Rev. X **3**, 021002 (2013).

25. L.P. Gor'kov, Proc. Natl. Acad. Sci. **113**, 4646 (2016).

26. J.M. Edge, Y. Kedem, U. Aschauer, N.A. Spaldin and A.V. Balatsky, Phys. Rev. Lett. **115**, 247002 (2015).

27. J. Ruhman and P.A. Lee, Phys. Rev. B **94**, 224515 (2016).

28. S.E. Rowley, C. Enderlein, J. Ferreira de Oliveira, D.A. Tompsett, E. Baggio Saitovitch, S.S. Saxena and G.G. Lonzarich, arXiv:1801.08121 (2018).

29. T.V. Trevisan, M. Schütt, and R.M. Fernandes, Phys. Rev. Lett. **121**, 127002 (2018).

30. T. Okuda, K. Nakanishi, S. Miyasaka and Y. Tokura, Phys. Rev. B **63**, 113104 (2001).

31. J. Fukuyado, K. Narikiyo, M. Akaki, H. Kuwahara and T. Okuda, Phys. Rev. B. **85**, 075112 (2012).

32. X. Lin, B. Fauque and K. Behnia, Science **349**, 945 (2015).

33. E. Mikheev, S. Raghavan, J.Y. Zhang, P.B. Marshall, A.P. Kajdos, L. Balents and S. Stemmer, Sci. Rep. **6**, 20865 (2016).

34. S. Stemmer and S.J. Allen, Rep. Prog. Phys. **81**, 062502 (2018).

35. See Supplemental Material at LINK for additional information on: prior literature heat capacity parameters; materials and methods; comparison of effective masses from heat capacity, SdH, and ARPES; $T^2$ resistivity analysis; Kadowaki-Woods scaling.

36. E. Ambler, J.H. Colwell, W.R. Hosler and J.F. Schooley, Phys. Rev. **148**, 280 (1966).

37. J.H. Colwell, Phys. Lett. **25A**, 623 (1967).

38. N.E. Phillips, B.B. Triplett, R.D. Clear, H.E. Simon, J.K. Hulm, C.K. Jones and R. Mazelsky, Physica **55**, 571 (1971).

39. I. Henning and E. Hegenbarth, Phys. Stat. Sol. (a) **83**, K23 (1984).

40. M. Ahrens, R. Merkle, B. Rahmati and J. Maier, Physica B **393**, 239 (2007).

41. A. Duran, F. Morales, L. Fuentes and J.M. Siqueiros, J. Phys.: Cond. Mat. **20**, 085219 (2008).

42. X. Lin, A. Gourgout, G. Bridoux, F. Jomard, A. Pourret, B. Fauque, D. Aoki and K. Behnia, Phys. Rev. B **90**, 140508(R) (2014).

43. M.J. Rice, Phys. Rev. Lett. **20**, 1439 (1968).

44. K. Kadowaki and S.B. Woods, Sol. Stat. Commun. **58**, 507 (1986).

45. W.D. Rice, P. Ambwani, M. Bombeck, J.D. Thompson, G. Haugstad, C. Leighton and S.A. Crooker, Nat. Mater. **13**, 481 (2014).

46. J.C. Lashley, M.F. Hundley, A. Migliori, J.L. Sarrao, P.G. Pagliuso, T.W. Darling, M. Jaime, J.C. Cooley, W.L. Hults, L. Morales, D.J. Thoma, J.L. Smith, J. Boerio-Goates, B.F. Woodfield, G.R. Stewart, R.A. Fisher, and N.E. Phillips, Cryogenics **43**, 369 (2003).

47. C.W. Rischau, X. Lin, C.P. Grams, D. Fick, S. Harms, J. Engelmayer, T. Lorenz, Y. Gallais, B. Fauque, J. Hemberger and K. Behnia, Nat. Phys. **13**, 643 (2017).

48. E.S.R. Gopal, *Specific Heats at Low Temperatures*, Plenum Press, New York (1966).

49. N. Choudhury, E.J. Walter, A.I. Kolesnikov and C.-K. Loong, Phys. Rev. B. **77**, 134111 (2008).

50. X. Lin, C.W. Rischau, L. Buchauer, A. Jaoui, B. Fauque and K. Behnia, npj Quantum Mater. **2**, 41 (2017).

51. Y. Tokura, Y. Taguchi, Y. Okada, Y. Fujishima, T. Arima, K. Kumagai and Y. Iye, Phys. Rev. Lett. **70**, 2126 (1993).

52. A.C. Jacko, J.O. Fjaerestad and B.J. Powell, Nat. Phys. **5**, 422 (2009).

53. N.E. Hussey, J. Phys. Soc. Jpn. **74**, 1107 (2005).

**Figure Captions**

**Figure 1:** (a) Specific heat ($C_P$) vs. temperature ($T$) from 1.8-280 K, with Nb contents ($x$) and Hall densities ($n$) labeled. The horizontal line marks the Dulong-Petit value. (b) $C_P/T$ vs. $T^2$ up to ~9 K, with solid line fits discussed in the text. Inset: Expanded view from 1.8-3.2 K. (c) $C_P/T^3$ vs. $T$ up to 80 K. The dashed line shows Debye $C_P(T)$ for a Debye temperature of 560 K; the solid line fit adds the Einstein contribution discussed in the text. (d) *Theoretical* electronic heat capacity, plotted as $C_V^{elec}/T$ vs. $T$ for the same $n$ values studied experimentally. Horizontal dashed lines mark the Sommerfeld coefficient as $T \rightarrow 0$.

**Figure 2:** Nb content ($x$) dependence of: (a) the Debye temperature ($\theta_D$) and (b) the $T^5$ specific heat prefactor ($\alpha$). (c) Hall electron density ($n$) dependence of the Sommerfeld coefficient ($\gamma$). Shown are experimental points, and theoretical values (from Fig. 1(d)) multiplied by 2.0. Dashed lines are guides to the eye.

**Figure 3:** (a) Band-structure from a tight-binding fit to a first-principles calculation [21]. Labeled are the three bands, the tetragonality and spin-orbit splittings, and the electron densities at the Lifshitz transitions. Energies are divided by the factor 2.0. (b) Density-of-states effective mass ($m^*_{DOS}/m_e$) vs. Hall density ($n$). Shown are experimental values from specific heat (black circles), theoretical values from the band-structure in (a) (red circles), approximate Shubnikov-de Haas values for the three bands (blue, red, green dashed lines, measured in the cubic [100] direction [16]), and the single $m^*_{DOS}$ from ref. 42. The black dashed line is a guide to the eye. (c) Kadowaki-Woods plot (log $A$ vs. log $\gamma^2$), where $A$ is the $T^2$ resistivity prefactor, and $\gamma$ is the Sommerfeld coefficient. Shown are Nb-doped SrTiO$_3$ (STO:Nb) results from this work (black open circles), La-doped SrTiO$_3$ (STO:La) results from Okuda *et al.* [30] (green diamonds), Sr-

doped LaTiO$_3$ (LTO:Sr) results from Tokura *et al.* [52] (orange squares), along with data on various transition metals, heavy fermion compounds, and oxides from [51]. Dashed lines are linear fits; the solid line is a guide to the eye.

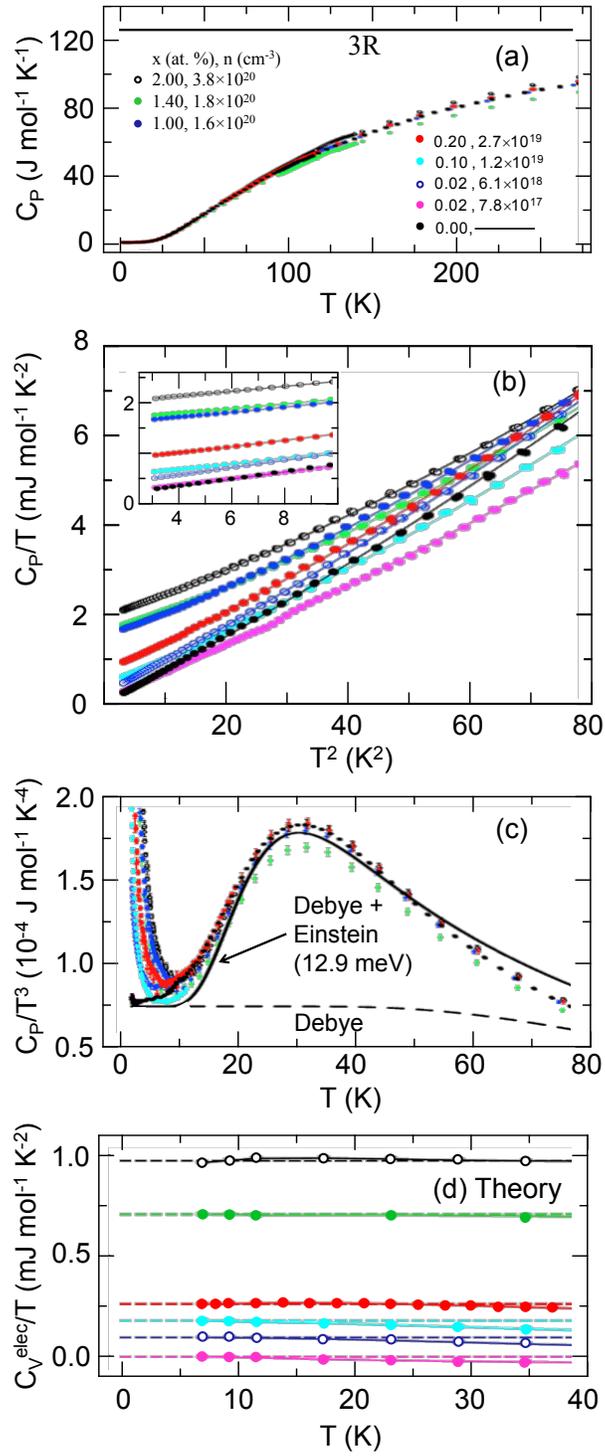

**Figure 1**

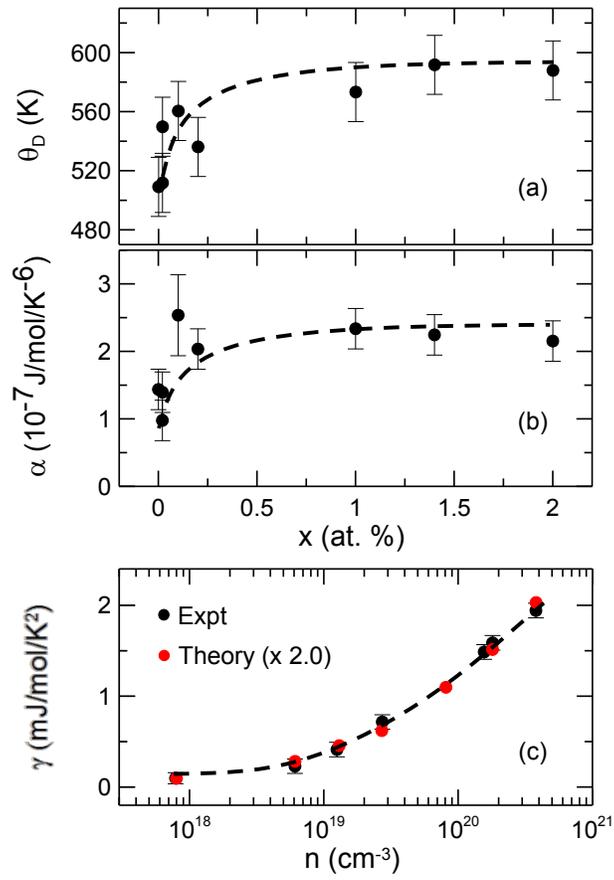

**Figure 2**

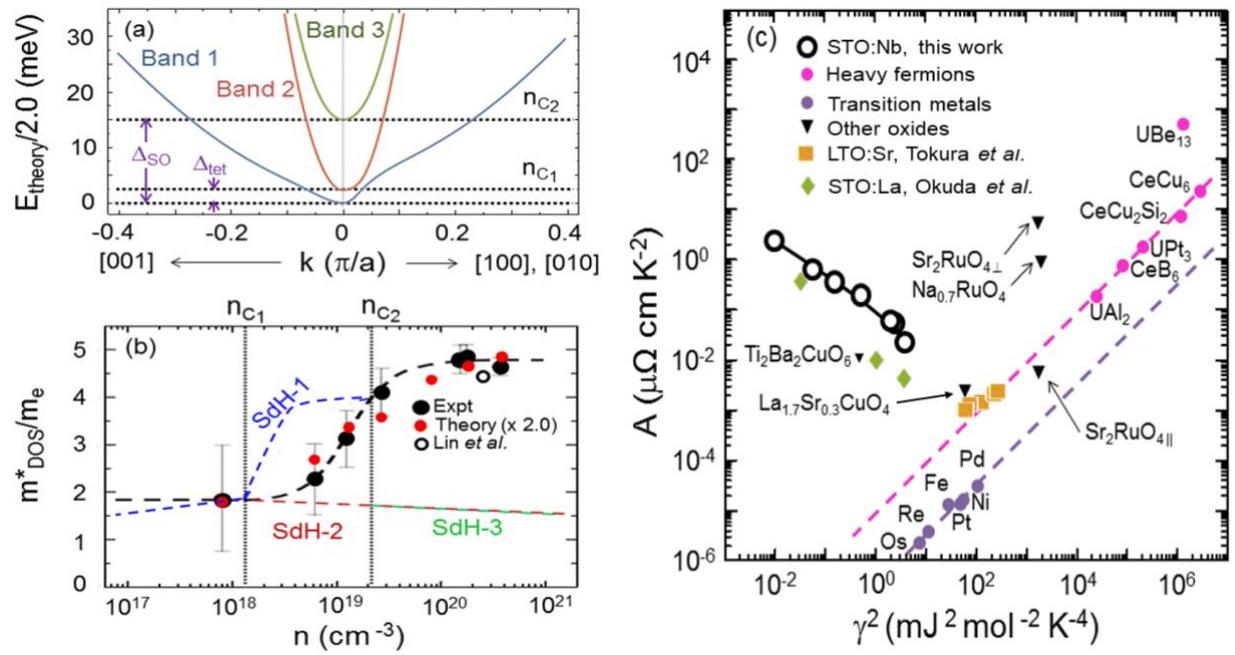

**Figure 3**

Supplemental Material for:

# Low Temperature Specific Heat of Doped SrTiO$_3$:
# Doping Dependence of the Effective Mass and Kadowaki-Woods Scaling Violation

E. McCalla[1,2], M. N. Gastiasoro[3], G. Cassuto[1], R.M. Fernandes[3] and C. Leighton[1]

[1]*Department of Chemical Engineering and Materials Science, University of Minnesota, Minneapolis, Minnesota, USA*

[2] *Department of Chemistry, McGill University, Montreal, Canada*

[3]*School of Physics and Astronomy, University of Minnesota, Minneapolis, Minnesota, USA*

**Section A: Prior Heat Capacity Results on Doped SrTiO$_3$ (Debye Temperatures and Density-of-States Effective Masses)**

As alluded to in the main text, prior literature reports on parameters derived from heat capacity measurements on SrTiO$_3$ are remarkably scattered. Table SI below shows the reported Debye temperatures and density-of-states effective masses for SrTiO$_3$, in undoped form, and when substituted/doped with either Nb, La, Ba, Ca, Pr, or oxygen vacancies (V$_O$). The values given here are specifically from low temperature analyses.

| Citation, year | Dopant/ $n$ (10$^{19}$cm$^{-3}$) | S or P | $\theta_D$ (K) | $m^*_{DOS}/m_e$ |
|---|---|---|---|---|
| Ahrens *et al.*, 2007 [S1] | Undoped | S | 513 | --- |
|  | V$_O$ / 6.0 | S | 513 | 1.8 |
|  | Nb / 5% | P | 534 | 1.9 |
| Phillips *et al.*, 1971 [S2] | V$_O$ / 3.6 | P | 303 | 5.2 |
|  | V$_O$ / 8.4 | P | 365 | 5.5 |
|  | V$_O$ / 11.6 | P | 235 | 7.8 |
|  | Nb / 3.1 | S | 435 | 5.7 |
| Colwell, 1967 [S3] | Ba (V$_O$?) / 9.6 | P | 453 | 4.72 |
| Ambler *et al.*, 1966 [S4] | Nb / 14 | P | 453 | 5.3 |
| Okuda *et al.*, 2001 [S5] | V$_O$ / 8.8 | S | 402 | 1.2 |
|  | La / 37.3 | S | 378 | 1.5 |
|  | La / 102.3 | S | 380 | 1.6 |
| Lin *et al.*, 2014 [S6] | Nb / 26 | S | --- | 4.2 |
| Fukuyado *et al.*, 2012 [S7] | Nb / 17.1 | S | 327 | 2.32 |
|  | Nb / 54.8 | S | 304 | 2.30 |
|  | Nb,Ca / 44.4 | S | 349 | 2.33 |
| Henning *et al.*, 1984 [S8] | Undoped | S | 324 | --- |
| Duran *et al.*, 2008 [S9] | Undoped | P | 428 | --- |
|  | Pr / 15% | P | 385 | --- |

**Table SI:** Literature Debye temperatures ($\theta_D$) and density-of-states effective masses ($m^*_{DOS}$) for SrTiO$_3$ (single crystal, S, or polycrystalline, P), either undoped, or doped/substituted with Nb, La, Ba, Ca, Pr or oxygen vacancies (V$_O$). $n$ is the reported Hall electron density. If no such density is available, a dopant concentration (at. %) is given.

As can be seen from the table, the scatter in the Debye temperature, $\theta_D$, is very wide, spanning from 235 to 534 K. Even in undoped SrTiO$_3$, the scatter spans from 324 to 513 K, revealing no clear consensus. This remains true even if considering only single crystals. Similar scatter is seen in the density-of-states effective mass, $m^*_{DOS}$. This is shown more clearly in Fig. S1 below, which plots $m^*_{DOS}(n)$ using the values shown in Table SI. The results scatter between 1.2 and 7.8$m_e$, with no clear trend with $n$. The literature data are also concentrated only above about $3 \times 10^{19}$ cm$^{-3}$. As discussed in the main text, our own results extend to much lower $n$, and reveal a systematic trend.

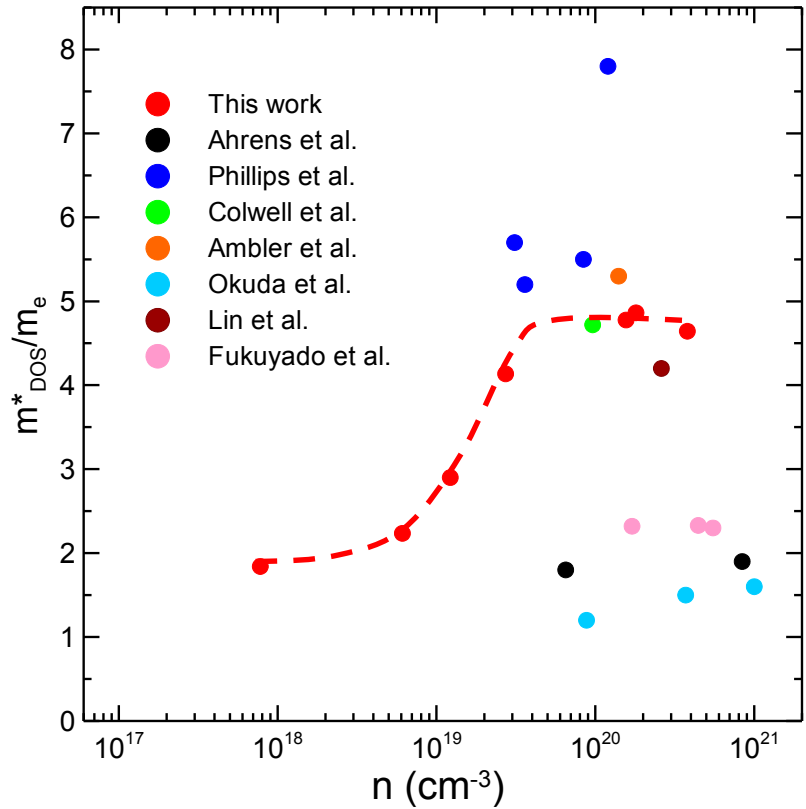

**Fig. S1**: Doping dependence of the density-of-states effective mass ($m^*_{DOS}/m_e$), compiling our own data, discussed in the main text, along with prior literature reports (from Table SI).

## Section B: Materials and Methods Details

Commercial single crystals of $SrTi_{1-x}Nb_xO_3$ with $x$ = 0, 0.02, 0.1, 0.2, 1.0, 1.4, and 2.0 at. %, from three suppliers (Crystec, Crystal GmbH, and MTI Corporation), were used in this study. Details on composition, supplier, 300 K Hall electron density ($n$), 300 K resistivity ($\rho$), and 4 K mobility ($\mu$) are shown in Table SII, along with some heat capacity parameters discussed below. As noted in the main text, some of these crystals were previously used in earlier studies by some of us [S10,S11]. They have been characterized by high-resolution X-ray diffraction [S10], temperature-dependent transport and magnetotransport measurements [S10], and trace impurity analysis [S11]. Note that substantial differences in $n$ are seen between the two crystals with 0.02 at. % Nb, likely due to differing compensation levels.

| $x$ (at. %) | Supplier | $n$(300 K) (cm$^{-3}$) | $\rho$(300 K) ($\Omega$ cm) | $\mu$(4 K) (cm$^2$ V$^{-1}$ s$^{-1}$) | $C_s/C_a$ (1.8 to 10 K) | Coupling (%) |
|---|---|---|---|---|---|---|
| 0.00 | Crystec | ---- | ---- | ---- | 0.17 → 0.37 | 65** |
| 0.02* | Crystal | $7.8 \times 10^{17}$ | 1.23 | 22100 | 0.23 → 0.35 | ----- |
| 0.02 | Crystec | $6.1 \times 10^{18}$ | 0.152 | 10100 | 0.35 → 0.50 | 94 |
| 0.10 | Crystec | $1.2 \times 10^{19}$ | 0.0798 | 4020 | 0.33 → 0.65 | 95 |
| 0.20* | Crystal | $2.7 \times 10^{19}$ | 0.0343 | 2370 | 0.27 → 0.80 | 95 |
| 1.00* | MTI | $1.6 \times 10^{20}$ | 0.00570 | 469 | 0.40 → 1.60 | 98 |
| 1.40* | MTI | $1.8 \times 10^{20}$ | 0.00690 | 570 | 0.71 → 3.00 | 98 |
| 2.00* | Crystal | $3.8 \times 10^{20}$ | 0.00241 | 316 | 0.51 → 2.50 | 98 |

**Table SII:** Doping level, supplier, 300 K Hall electron density ($n$), 300 K resistivity ($\rho$), 4 K mobility ($\mu$), ratio of sample to addenda heat capacities ($C_s/C_a$), and thermal coupling percentages for the Nb-doped $SrTiO_3$ bulk single crystals studied in this work. The $C_s/C_a$ values are for temperatures in the range 1.8 to 10 K. The thermal coupling percentages shown are the *minimum* values obtained; no entry indicates a failure of the two τ fitting method, as discussed below.

* Transport data first reported in ref. [S10]

** Value was above 90% for $T > 4.7$ K. $C_s/C_a$ was above 0.20 for $T > 3.0$ K.

Specific heat ($C_p$) measurements were made between 1.8 and 280 K in a Quantum Design Physical Property Measurement System (PPMS), using short pulse (2% of the measurement temperature) relaxation calorimetry. In all cases the addenda heat capacity ($C_a$) was first determined, measuring only the sample platform and Apiezon N grease used to affix the samples. Crystals were then added and the sample heat capacity ($C_s$) determined by subtraction of addenda. Two quality parameters were continuously monitored: The PPMS sample-calorimeter coupling percentage, and $C_s/C_a$. Coupling percentages above 90% are recommended for minimal systematic errors. Lashley *et al*. [S12] have recommended minimum $C_s/C_a$ ratios for specific systematic uncertainties on $C_P$. They suggest that at $C_s/C_a = 0.5$ one can expect 3% error, at $C_s/C_a = 0.2$, 5% error, and at $C_s/C_a = 0.2$, poor accuracy, with errors up to 30%. This is a particular issue at low temperatures when the Sommerfeld coefficient is small or negligible, due to the low total $C_s$.

Minimum values of thermal coupling percentage, as well as the relevant range of $C_s/C_a$ values (between 1.8 and 10 K) are provided for our samples in Table SII. It can be seen that $C_s/C_a$ is maintained at 0.2 and above in all cases (*i.e.*, <5% errors), with the exception of the very lowest temperatures (<3 K) in the undoped crystal. Similarly, the coupling was excellent in all cases, for all crystals, except the undoped and $n = 7.8 \times 10^{17}$ cm$^{-3}$ crystals. For the undoped crystal, the coupling dipped below 90% only below 4.7 K. For the $n = 7.8 \times 10^{17}$ cm$^{-3}$ crystal, significant coupling problems occurred around 5 K, where non-negligible fluctuations begin to emerge in $C_P(T)$ in Fig. 1(b). It must be emphasized, however, that the fit range in Fig. 1(b) from which our key parameters are extracted extends significantly above the temperature range where these problems arise. Additionally, and as emphasized in the main text, we obtain excellent agreement between Debye temperatures and Sommerfeld coefficients obtained from the fits in Fig. 1(b) (to 9 K), and its inset (to 3-4 K), generating yet further confidence.

For the crystals not previously studied in ref. [S10], resistivity and Hall effect measurements were made between 4 and 300 K in a variety of cryostats/magnets. Indium contacts were employed, in a van der Pauw geometry, using AC excitation. Extensive checks were made for non-Ohmicity and self-heating.

**Section C: Details on Comparing Effective Masses from Heat Capacity with Shubnikov de Haas (SdH) measurements**

As discussed in the main text, electronic specific heat provides information on the density-of-states effective mass, $m^*_{DOS}$, which can be compared to the transport effective mass from Shubnikov de Haas oscillations, $m^*$. To do this we define $m^*_i$ and $n_i$ as the SdH mass and electron density in each band, $i$. If each band $i$ is approximated by a parabolic dispersion, the Sommerfeld coefficient is given by $\gamma(n) = \sum_{i=1}^{3} n_i^{1/3} m_i^* (k_B/\hbar)^2 (\pi/3)^{2/3}$. Assuming an effective parabolic single-band for the entire multi-band system, we can also write $\gamma(n) = m^*_{DOS}(n) n^{1/3} (k_B/\hbar)^2 (\pi/3)^{2/3}$. Thus, within these approximations, $m^*_{DOS}$ from heat capacity measurements can be related to the transport effective masses $m^*_i$ by $m^*_{DOS} = \sum_{i=1}^{3} n_i^{1/3} m_i^* / (\sum_{i=1}^{3} n_i)^{1/3}$. Using this approach, a comparison of the doping-dependent density-of-states effective mass from heat capacity and the transport effective mass from SdH is provided in the main text, in connection with the data in Fig. 3(b). Additional details on the calculations are presented below, in the three important filling regimes.

$n < n_{c1}$: Here, the single occupied band is approximated by a parabolic dispersion centered at the Γ point in a tetragonal system, i.e., $E(k) = \frac{\hbar^2}{2m^*_p}(k_x^2 + k_y^2) + \frac{\hbar^2}{2m^*_z} k_z^2$ with degenerate in-plane mass $m^*_p \equiv m^*_x = m^*_y$. The corresponding effective density-of-states mass is given by $m^*_{DOS} = \sqrt[3]{m*_p^2\, m*_z}$. The experimental $m^*_{DOS}$ can thus be compared with $m^*_{DOS} = \sqrt[3]{m*_p^2\, m_z}$, with transport effective masses extracted from the band structure deduced from SdH oscillations in ref. [S13]. At $n = n_{c1}$, $m^*_p = 3.6$ and $m^*_z = 1.68$, giving a value $m^*_{DOS} = 2.8 m_e$. Note that the SdH values plotted in Fig. 3(b) in the main text are from a study [S14] performed in only one orientation (cubic [100]). Those data are shown here due to the completeness of the $n$-dependent data set, despite the lack of angle dependence.

$n_{c1} < n < n_{c2}$: Here, two bands are occupied. Their transport effective masses are again extracted from ref. [S13]. At $n_{c2}$, for band 1, this gives: $m_{1p} = 5.45 m_e$ and $m_{1z} = 1.68 m_e$, resulting in an effective mass for band 1 of $m_1^* = \sqrt[3]{m*_{1p}^2\, m_{1z}} = 3.68 m_e$. Similarly, for band 2 at $n_{c2}$, $m_p = 1.59 m_e$ and $m_z = 4.92 m_e$, yielding $m_2^* = 2.32 m_e$. The values of $n_i$ can then be obtained from SdH oscillations from ref. [S14]. Extrapolation to $n = n_{c2}$ gives $n_1 = 1.2 \times 10^{19}$ cm$^{-3}$ and $n_2 = 4.5 \times 10^{18}$ cm$^{-3}$, giving a total $n = 1.65 \times 10^{19}$ cm$^{-3}$ (*i.e.*, the sum of SdH $n_i$ values is below the actual $n$, as noted by Lin *et al.* [S14]). With these SdH mass values, we then obtain $m^*_{DOS} = 4.9m_e$ at $n_{c2}$, as quoted in the text. $m^*_{DOS}$ is thus predicted to gradually grow to $4.9m_e$ at $n_{c2}$, in reasonable agreement with our data from heat capacity (see Fig. 3(b)).

***n > n*$_{c2}$:** As pointed out in the main text, in this regime SdH data become sparse, especially for band 1, due to low mobility. Data are available at $n = 1.5 \times 10^{20}$ cm$^{-3}$, however, and we thus focus on this doping level. All three bands are now occupied, with estimated occupations from ref. [S14] of $n_1 = 8.0 \times 10^{19}$ cm$^{-3}$ (extrapolated), $n_2 = 1.6 \times 10^{19}$ cm$^{-3}$ and $n_3 = 6.0 \times 10^{17}$ cm$^{-3}$, giving $n = 9.7 \times 10^{19}$ cm$^{-3}$, again slightly lower than the measured total $n$, as noted by Lin *et al.* From ref. S14, the masses for each of the upper bands are available in the $z$-direction only; assuming isotropic behavior gives $m^*_2 = 1.6m_e$ and $m^*_3 = 1.5m_e$. For the lowest band, ARPES gives an effective mass of $7.0m_e$. This band is known to be anisotropic, however, so we use $m_x = m_y = 7.0m_e$ and $m_z = 1.68m_e$ (the SdH value at the highest studied concentration $n_{c2}$). We consider this a reasonable estimate based on the band structure of Uwe *et al.*, yielding $m^*_1 = 4.35m_e$. This then yields a predicted $m^*_{DOS} = 5.2m_e$, in good agreement with our heat capacity values in Fig. 3(b).

**Section D: $T^2$ Resistivity Analysis**

Shown below in Fig. S2 are illustrative resistivity ($\rho$) *vs.* $T^2$ plots for four doped crystals, in the $10^{17}$, $10^{18}$, $10^{19}$, and $10^{20}$ cm$^{-3}$ electron density ranges. Included are straight line fits in the best-fit $T^2$ region, which is below 50-110 K, weakly dependent on doping. We emphasize that deviations to a weaker temperature dependence than $T^2$ can be seen at the lowest $T$ in some cases, particularly at the extremes of $n$ studied. More detailed analysis is provided in Fig. S3, which plots the temperature dependence of the exponent, $m$, in $\rho = \rho_0 + AT^m$, for the same four crystals. This was determined by numerical differentiation of the data, smoothed with adjacent point averaging over 5 data points (9 data points for the highest $n$ sample, where the $T$ dependence is weak). At the lowest doping, $7.8 \times 10^{17}$ cm$^{-3}$, $m$ first increases on cooling from 300 K, hitting a maximum at ~3.5 before approaching 2.0 at low temperatures. At $6.1 \times 10^{18}$ and $1.2 \times 10^{19}$ cm$^{-3}$, however, $m$ indeed becomes very close to 2.0 over a significant temperature window, consistent with the very good fits in Fig. S2(b,c). The same overall shape of $m(T)$ is seen as for the lightest doped sample, the peak being at $m \approx 3$ in these cases. Finally, at $3.8 \times 10^{20}$ cm$^{-3}$, $m \approx 2$ is seen up to as high as 110 K, above which it drifts towards 3. While some limitations are apparent, as noted above, we thus conclude, in agreement with prior work, that $\rho = \rho_0 + AT^2$ describes well the behavior in doped SrTiO$_3$ over a significant temperature window.

Fig. S4 below shows the resulting $A$ values plotted *vs.* the Hall electron density, $n$. The results from the seven doped samples studied in this work are shown by the open black circles, superimposed on the earlier results from Okuda *et al.*, [S5], Lin *et al.* [S15] and Mikheev *et al.* [S16]. The overall agreement is good. As previously noted, $A \propto 1/n$ is followed quite closely over a four order of magnitude range in doping. As emphasized by Mikheev *et al.*, $A \propto 1/n$ indicates an essentially doping-independent electron scattering rate.

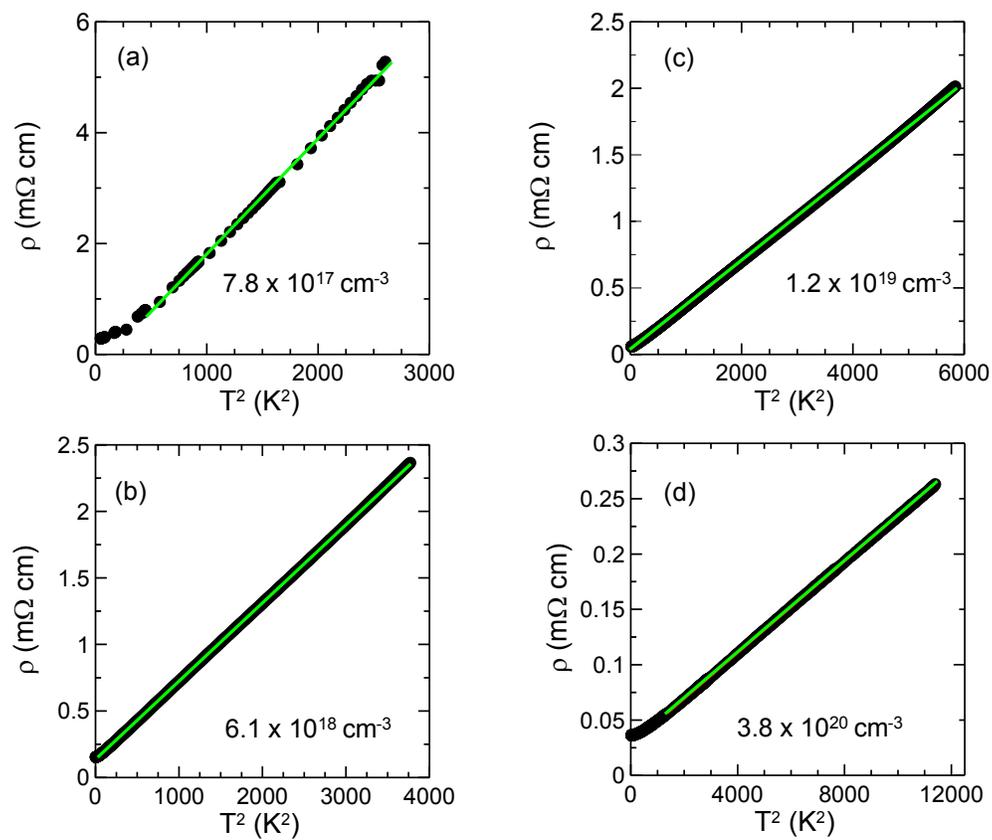

**Fig. S2:** Resistivity (ρ) *vs.* $T^2$ plots for four illustrative samples with Hall electron densities of $7.8 \times 10^{17}$, $6.1 \times 10^{18}$, $1.2 \times 10^{19}$, and $3.8 \times 10^{20}$ cm$^{-3}$. The green solid lines are straight line fits.

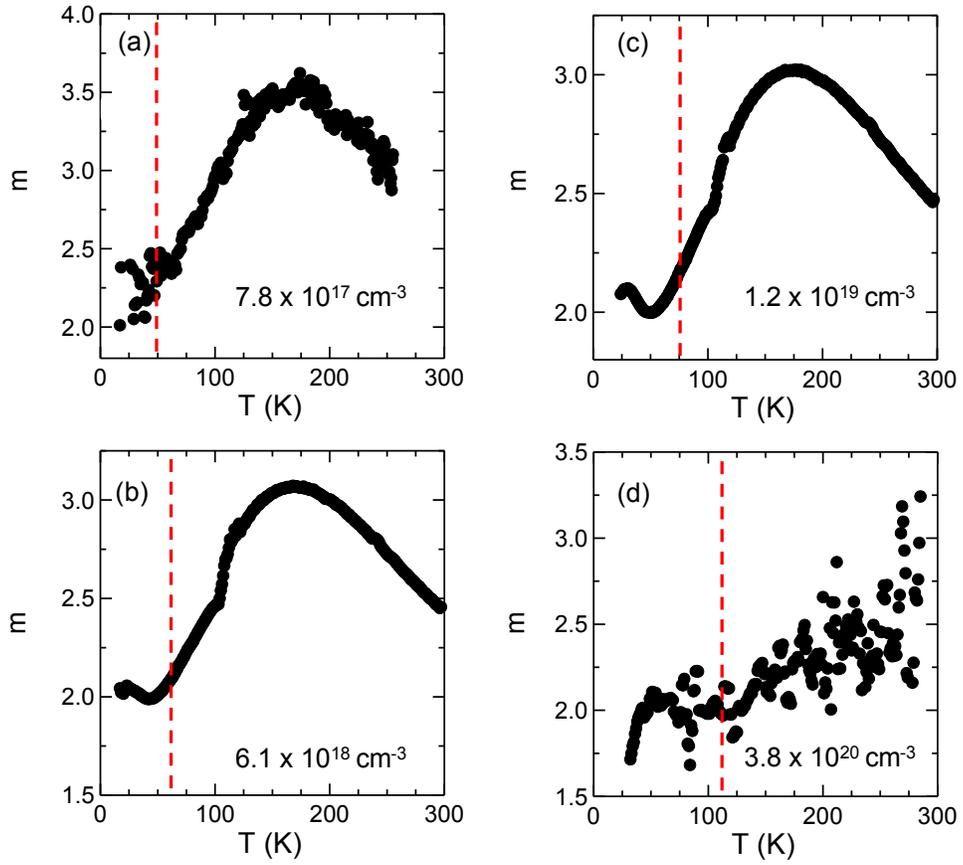

**Fig. S3:** Temperature dependence of the exponent, *m*, in $\rho = \rho_0 + AT^m$, for four illustrative samples with Hall electron densities of $7.8 \times 10^{17}$, $6.1 \times 10^{18}$, $1.2 \times 10^{19}$, and $3.8 \times 10^{20}$ cm$^{-3}$. The vertical dashed lines mark the highest temperature used for the $T^2$ fit in Fig. S2.

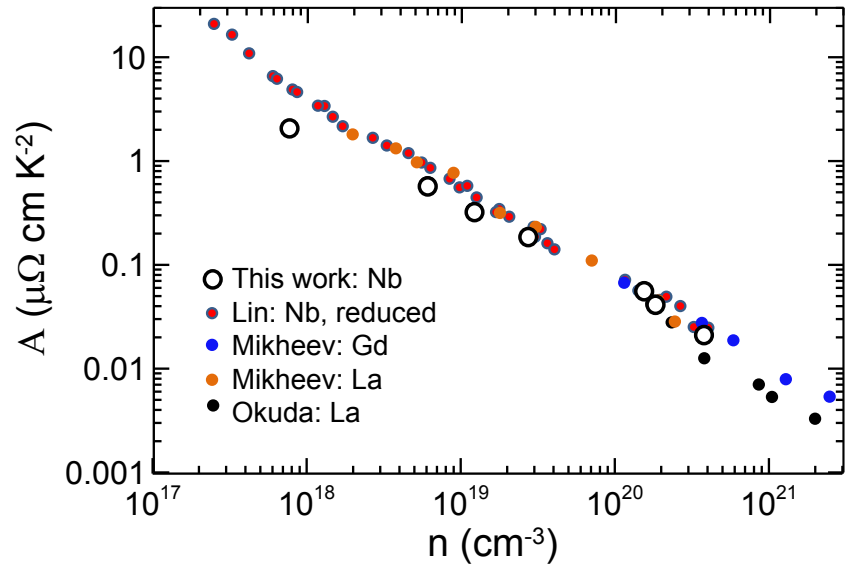

**Fig. S4:** Hall electron density ($n$) dependence of the $T^2$ resistivity prefactor ($A$) from this work (open circles) and the prior studies of Okuda *et al.*, [S5], Lin *et al.* [S15] and Mikheev *et al.* [S16]. The relevant citation and dopant are given in the legend.

**Section E: Modified Kadowaki-Woods Scaling**

Jacko *et al.* [S17] have proposed a modified form of Kadowaki-Woods scaling that takes into account carrier density and spatial dimensionality. The solid line in Fig. S5 below shows the scaling thus obtained for transition metals, organic conductors, heavy fermion compounds, and oxides. Rather than $A$ vs. $\gamma^2$, the plot shows $A$ vs. $\gamma^2/f_{dx}$, where $d$ is the dimensionality (3 in our case), and $f_{dx}(n) = nD_0^2\langle v_{0x}^2\rangle\xi^2$. Here, $n$ is the conduction electron density, $D_0$ is the bare density-of-states at the Fermi energy, $\langle v_{0x}^2\rangle$ is an average over the fermi surface of the $x$-component of the squared Fermi velocity, and $\xi \approx 1$. For an isotropic Fermi liquid this reduces to $f_{3x}(n) \approx (3n^7/\pi^4 \hbar^6)^{1/3}$. As shown below, while this parameterization does generate $A$ that *increases* with $\gamma^2/f_{dx}$, collapse to the solid black line still does not occur. Thus, even the modified Kadowaki-Woods scaling of Jacko *et al.* is violated in SrTiO$_3$.

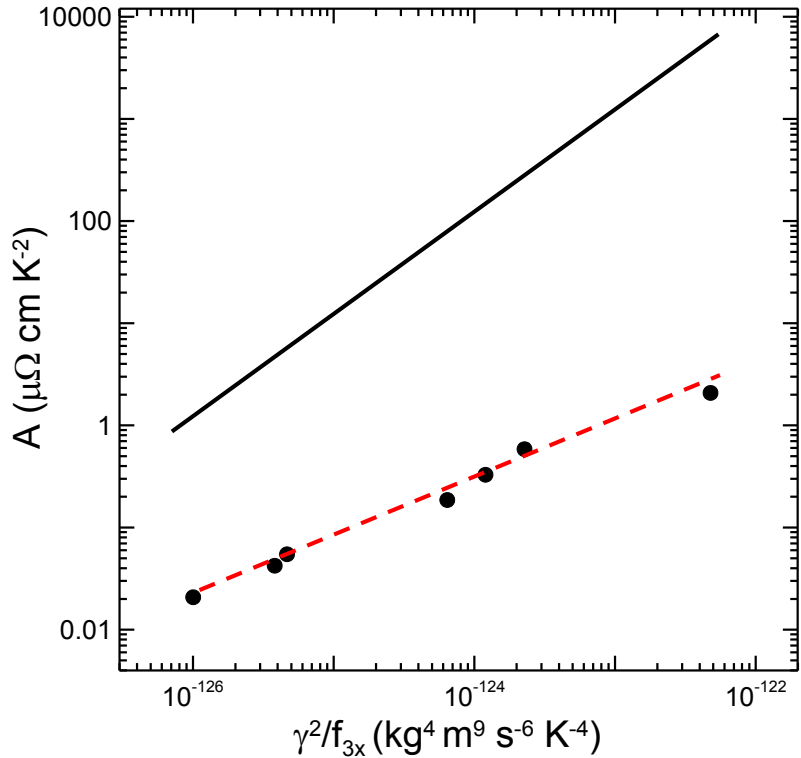

**Fig. S5:** Modified Kadowaki-Woods scaling plot of the type proposed by Jacko *et al.* [S]. The solid black line is the line to which transition metals, heavy fermion compounds, organic conductors, and oxides collapse. The solid black points and dashed red fit line are for our SrTiO$_3$ crystals.

Hussey [S18] has also proposed a modified approach to Kawdowaki-Woods scaling to account for carrier density effects. This was shown to be successful in the $Sr_{1-x}La_xTiO_3$ system [S18]. The method involves expressing the Sommerfeld coefficient, $\gamma$, in its volume form $\gamma_V = \frac{\gamma}{N_A V}$ where the unit cell volume $V = a^3$, the lattice spacing $a = 3.905$Å in this case, and $N_A$ is the Avogadro number. The effect of carrier density, $n$, in spherical 3D systems is then included via the Fermi wavevector, and the proposed rescaled Kadowaki-Woods ratio becomes $\frac{A}{\gamma_V^2} \propto C \frac{a}{n^2}$, where the parameter $C$ is shown in Fig. S6 below. As shown in the figure, our data on $SrTiO_3$:Nb are consistent with this form of scaling, in addition to those on the $Sr_{1-x}La_xTiO_3$ system. This form of scaling thus captures carrier density effects in both doped $SrTiO_3$ and doped $LaTiO_3$.

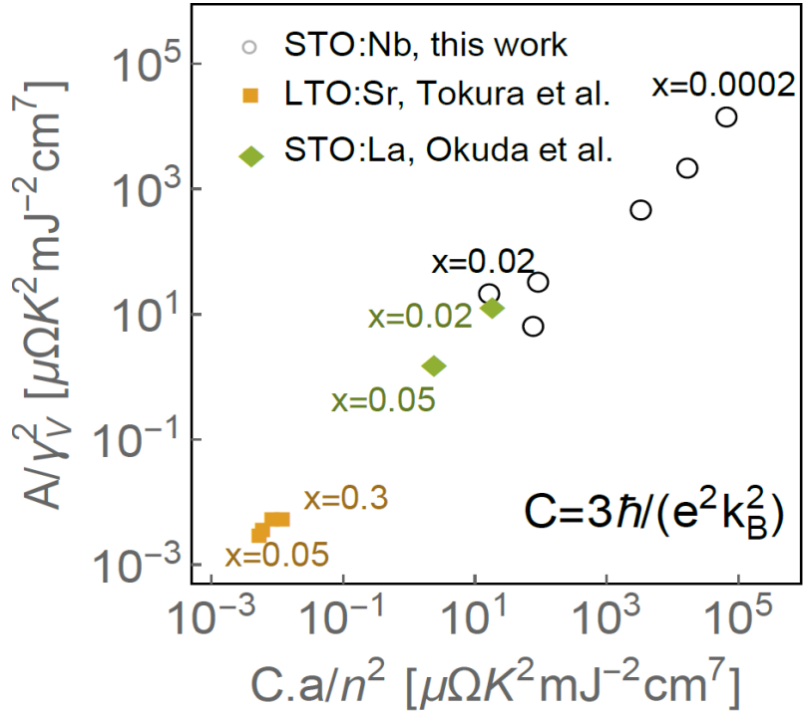

**Fig. S6:** Modified Kadowaki-Woods plot, following Hussey [S18]. Data are shown for $SrTiO_3$:La (STO:La) and $LaTiO_3$:Sr (LTO:Sr) [S18, originally from the Okuda *et al.* and Tokura *et al.* works cited in the main text], as well as our own $SrTiO_3$:Nb (STO:Nb). Symbols are defined in the paragraph above.

**Supplemental Material References**

[S1] M. Ahrens, R. Merkle, B. Rahmati and J. Maier, Physica B **393**, 239 (2007).

[S2] N.E. Phillips, B.B. Triplett, R.D. Clear, H.E. Simon, J.K. Hulm, C.K. Jones and R. Mazelsky, Physica **55**, 571 (1971).

[S3] J.H. Colwell, Phys. Lett. **25A**, 623 (1967).

[S4] E. Ambler, J.H. Colwell, W.R. Hosler and J.F. Schooley, Phys. Rev. **148**, 280 (1966).

[S5] T. Okuda, K. Nakanishi, S. Miyasaka and Y. Tokura, Phys. Rev. B **63**, 113104 (2001).

[S6] X. Lin, A. Gourgout, G. Bridoux, F. Jomard, A. Pourret, B. Fauque, D. Aoki and K. Behnia, Phys. Rev. B **90**, 140508(R) (2014).

[S7] J. Fukuyado, K. Narikiyo, M. Akaki, H. Kuwahara and T. Okuda, Phys. Rev. B. **85**, 075112 (2012).

[S8] I. Henning and E. Hegenbarth, Phys. Stat. Sol. (a) **83**, K23 (1984).

[S9] A. Duran, F. Morales, L. Fuentes and J.M. Siqueiros, J. Phys.: Cond. Mat. **20**, 085219 (2008).

[S10] A. Spinelli, M.A. Torija, C. Liu, C. Jan and C. Leighton, Phys. Rev. B. **81**, 155110 (2010).

[S11] W.D. Rice, P. Ambwani, M. Bombeck, J.D. Thompson, G. Haugstad, C. Leighton and S.A. Crooker, Nat. Mater. **13**, 481 (2014).

[S12] J.C. Lashley, M.F. Hundley, A. Migliori, J.L. Sarrao, P.G. Pagliuso, T.W. Darling, M. Jaime, J.C. Cooley, W.L. Hults, L. Morales, D.J. Thoma, J.L. Smith, J. Boerio-Goates, B.F. Woodfield, G.R. Stewart, R.A. Fisher, and N.E. Phillips, Cryogenics **43**, 369 (2003).

[S13] H. Uwe, R. Yoshizaki, T. Sakudo, A. Izumi and T. Uzumaki, Jpn. J. Appl. Phys. **24**, 335 (1985); H. Uwe, T. Sakudo and H. Yamaguchi, Jpn. J. Appl. Phys. **24**, 519 (1985).

[S14] X. Lin, G. Bridoux, A. Gourgout, G. Seyfarth, S. Kramer, M. Nardone, B. Fauque and K. Behnia, Phys. Rev. Lett. **112**, 207002 (2014).

[S15] X. Lin, B. Fauque and K. Behnia, Science **349**, 945 (2015).

[S16] E. Mikheev, S. Raghavan, J.Y. Zhang, P.B. Marshall, A.P. Kajdos, L. Balents and S. Stemmer, Sci. Rep. **6**, 20865 (2016).

[S17] A.C. Jacko, J.O. Fjaerestad and B.J. Powell, Nat. Phys. **5**, 422 (2009).

[S18] N.E. Hussey, J. Phys. Soc. Jpn. **74**, 1107 (2005).